\documentclass[12pt,usenatbib]{mn2e}
\usepackage{graphicx}
\usepackage{epsf}
\usepackage{color}
\usepackage{rotating}
\def\lsim{\mathrel{\lower0.6ex\hbox{$\buildrel {\textstyle <}
 \over {\scriptstyle \sim}$}}}
\def\gsim{\mathrel{\lower0.6ex\hbox{$\buildrel {\textstyle >}
 \over {\scriptstyle \sim}$}}}

\def\rvir{r_{\rm vir}}

\begin{document}

\title[Dark matter and Stellar haloes]{Disentangling the dark matter halo from the stellar halo}
\author[Libeskind et al.] 
{Noam I Libeskind$^1$, Alexander Knebe$^2$, Yehuda Hoffman$^3$,  Stefan
      Gottl\"ober$^1$,  \newauthor Gustavo Yepes$^2$\\
%   {Libeskind et alibi$^1$\\
          $^1$Leibniz-Institute f\"ur Astrophysik Potsdam, An der Sternwarte 16, D-14482 Potsdam, Germany\\
  $^2$Grupo de Astrof\'\i sica, Departamento de Fisica Teorica, Modulo C-XI, Universidad Aut\'onoma de Madrid, Cantoblanco E-280049, Spain\\
  $^3$Racah Institute of Physics, The Hebrew University of Jerusalem, Givat Ram, 91904, Israel
  }
%\date{Accepted 1988 December 15. Received 1988 December 14; in original form 1988 October 11}

%\pagerange{\pageref{firstpage}--\pageref{lastpage}} \pubyear{2002}

\maketitle \begin{abstract} \vspace{1pt}

The outer haloes of the Milky Way (MW) and Andromeda (M31) galaxies contain as much important information on their assembly and formation history as the properties of the discs resident in their centres.  Whereas the structure of dark matter (DM) haloes has been studied for a long time, new observations of faint structures hiding in the depths of the stellar halo have opened up the question of how the stellar halo is related to the DM underlying it. In this paper we have used the Constrained Local UniversE Simulation (CLUES) project to disentangle the stellar and DM component of three galaxies that resemble the MW, M31 and M33 using both DM only simulations and DM + gas-dynamical ones. We find that stars accreted in substructures and then stripped follow a completely different radial distribution than the stripped DM: the stellar halo is much more centrally concentrated than DM. In order to understand how the same physical process - tidal stripping  -  can lead to different $z=0$ radial profiles, we examined the potential at accretion of each stripped particle. We found that star particles sit at systematically higher potentials than DM, making them harder to strip. We then searched for a threshold in the potential of accreted particles $\phi_{\rm th}$, above which DM particles in a DM only simulation behave as star particles in the gas-dynamical one. We found that in order to reproduce the radial distribution of star particles, one must choose DM particles whose potential at accretion is $\gsim16\phi_{\rm subhalo}$, where $\phi_{\rm subhalo}$ is the potential at a subhaloes edge at the time of accretion.  A rule as simple as selecting particles according to their potential at accretion is able to reproduce the effect that the complicated physics of star formation has on the stellar distribution.  
This result is universal for the three haloes studied here and reproduces the stellar halo to an accuracy of within $\sim2\%$. Studies which make use of DM particles as a proxy for stars will undoubtedly miscalculate their proper radial distribution and structure unless particles are selected according to their potential at accretion. Furthermore, we have examined the time it takes to strip a given star or DM particle after accretion. We find that, owing to their higher binding energies, stars take longer  to be stripped than DM. The stripped DM halo is thus considerably older than the stripped stellar halo.

\end{abstract}
%\keywords{galaxy formation: general --- Black Holes, galaxy formation}

\section{Introduction}
\label{introduction} 
According to the current accepted cosmological model, the universe is composed of 26\% cold dark matter (DM), 70\% dark energy ($\Lambda$) and 4\% baryons \citep[e.g.][]{2007ApJS..170..377S}. Structure in the so-called $\Lambda$CDM cosmology forms from the bottom up \citep{1978MNRAS.183..341W,1985ApJ...292..371D} - the first objects to collapse at high redshift are small sub-galactic units, that condense out of small perturbations in the initial density field. These clumps then merge in an hierarchical fashion to construct the large bound objects we observe in the local universe. Visible luminous matter -  stars - are believed to be formed when Giant Molecular Clouds collapse in the potential wells of these bound blobs of DM and gas.

The merging process gives rise to DM haloes, which today host bright central galaxies such as the Milky Way (MW) and the Andromeda galaxy (M31) in their cores. The outskirts of such DM haloes are populated by a two component medium: diffuse matter and matter bound to substructures. Much of the mass is found bound to satellite galaxies which orbit within their parent halo. The properties (age, orbital parameters, spatial distribution, kinematics, etc)  of luminous satellite galaxies can teach us a lot regarding the formation of their hosts and have been the target of numerous observational and theoretical studies. Indeed, the past 5 years has seen an increased focus on the detection of satellite galaxies and has resulted in around a dozen new satellites being detected by the SDSS \citep{2008ApJ...686..279K,2009MNRAS.397.1748B}. 

Yet the $z=0$ satellite galaxy population is not a full survey of all the substructures accreted by the parent DM halo, since many substructures accreted at high redshift will, by $z=0$, have been tidally disrupted by the host potential, resulting in the stripping of dark matters and stars. Indeed it is believed that the stellar halo - stars exterior to the central galaxy and not bound to substructures - was formed by the tearing of stars from accreted satellite galaxies. \cite{2010MNRAS.406..744C} argue that the vast majority of stars in the MW's halo were stripped from just one or two large satellites. \cite{2009ApJ...702.1058Z} have studied the stellar halo in gas-dynamical/N-body simulations and have identified that in fact the stellar halo has a dual-origin: part of it was created via tidal stripping of stars from disrupted satellites, and part was pushed out of central galaxies during minor mergers. In a follow up paper,  \cite{2010ApJ...721..738Z} argue that the metallic abundance patterns (of [Fe/H] and [O/Fe]) of stars can be used to distinguish between theses different formation mechanisms.

Observations using the SDSS by, e.g., \cite{2008ApJ...680..295B} have indicated that the MW's halo is consistent with being formed entirely out of accreted debris material. This is in disagreement with observations by \cite{2008Natur.451..216C} who find clear differences between kinematical properties of the inner and outer stellar halo - stars in the inner halo are found to exhibit a net prograde rotation while the outer halo is dominated by retrograde motion. It has been suggested that the differences in net rotation of stars in the halo betray a dual origin, a result recently supported by \cite{2011arXiv1104.2513B} who find differences in kinematics for inner and outer halo stars.

The situation with M31 is similar as observations seem to favor an accreted origin. By studying the ages of stars in the halo, \cite{2008ApJ...685L.121B} argue against a in-situ origin and, because the stars are by and large relatively old - they argue for an hierarchical build up of M31. By focusing on the metal enrichment, \cite{2009ApJ...701..776G} too seem to argue for an entirely accreted stellar halo with little evidence of in-situ star formation.

The assembly history of the DM halo on the other hand, is more difficult to pin down as direct observations are by definition impossible. Yet many authors \citep[e.g.][and references therein]{2001ApJ...554..903K,2002ApJ...568...52W,2003MNRAS.339...12Z,2007ApJ...667..859D} have used $N$-body simulations to determine the relative importance of the two main ``modes'' of halo growth: diffuse accretion versus mergers. Most recently  \cite{2010arXiv1008.5114W} have used the Aquarius simulation and found that ambient accretion contributes the largest amount of material to the dark halo.

It is thus unclear if the conclusions drawn from observations of our stellar halo - that the stars were stripped from infalling satellite galaxies - are consistent with DM only simulations which point towards a diffuse smoothly accreted halo where mergers and debris material play a minor role. In this paper we disentangle these two components in order to understand how they co-evolved.

\section{Methods}
\label{sec:methods}
In this section we describe in brief the simulations used as well as the halo and subhalo finding algorithm employed to identify satellites.

\subsection{Constrained Simulations of the Local Group}
The simulations used in this work are embedded in the Constrained Local UniversE Simulation (CLUES) project and have been already studied in a number of recent papers \citep[e.g][]{2010MNRAS.401.1889L,2010MNRAS.405.1119K,2010MNRAS.402.1899K,2011MNRAS.411.1525L,2011MNRAS.412..529K} We refer the reader to those papers \citep[in particular][]{2010MNRAS.401.1889L} for details on how the constraints were generated and how the simulations were run: we highlight just the salient points here for clarity.

We choose to run our simulations using standard $\Lambda$CDM initial conditions, that assume a Wilkinson Microwave Anisotropy Probe 3 cosmology \citep{2007ApJS..170..377S} , i.e. $\Omega_{\rm m} =0.24$, $\Omega_{\rm b} = 0.042$, $\Omega_{\Lambda} = 0.76$ and $h=0.73$. We use a normalization of 
$\sigma_{8} = 0.75$ and an $n=0.95$ slope of the power spectrum. We use the MPI code \textsc{gadget2} \citep{2005MNRAS.364.1105S} to simulate the evolution of a cosmological box with side length of $L_{\rm box} = 64 h^{-1} Mpc$, before applying the zoom technique \citep[see e.g.][]{2001ApJ...554..903K} around a region of interest.

Instead of seeding our initial conditions as a just a random cube of space, the initial conditions of our volume are constrained to reproduce, at $z=0$ a number of objects that compose the local environment \citep[see][for details on how constraints initial conditions are generated]{1991ApJ...380L...5H}, including a ``virgo'' cluster, a ``coma'' cluster and a ``Local Group''. Our method allows us to properly constrain the large scales (i.e. those still linear by $z=0$) but we do not constrain the local group itself. In order to obtain a local group in the correct environment, three low resolutions constrained simulations are run with varying random seeds. Each $z=0$ low resolution simulations is then examined and if an object that resembles the local group is found (by construction this will be in the correct place), these initial conditions are selected for high resolution re-simulation. 

Our initial density field includes both DM and gas particles. Under the right conditions, gas particles may spawn star particles which interact gravitationally in the same way as DM (i.e. as point particles with a given softening length). Each gas particle may have up to two star formation episodes, each time spawning a star of half its original mass. In order to conserve mass we reduce the gas particle's mass each time a star particle is spawned, resulting in gas particles that have one or two times the mass of star particles (corresponding to gas particles that have spawned one or no star particles). When a gas particle spawns its second star particle, it ceases to exits.  Star particles represent stellar populations and are given the metalicity of the gas particle that spawned it. The massive stars born (with $M > 10M_{\odot}$) in this population explode instantaneously as supernovae type II, polluting the environment with metals and producing stellar winds. More details on the star formation prescription can be found in \citep{2010MNRAS.401.1889L}.

We resimulate just the region of interest around the local group. We centre a sphere of radius $2~h^{-1} \rm Mpc$ around the local group and populate it with $\sim 5.2\times10^{7}$ low mass, high resolution particles. Within our local group we are thus able to achieve a particle mass of just $M_{\rm dm} = 2.54  \times 10^{5}~h^{-1}\rm M_{\odot}$ for DM and $M_{\rm star} = 2.21  \times 10^{4}~h^{-1}\rm M_{\odot}$ for star particles.

Our constraints reproduce a cosmography which closely resembles the observed Local Group. In Table.~\ref{table:cosm} we compare properties of the simulated local group with observations of the real one\footnote{In a future paper we intend to study in detail the cosmography produced by our constrained simulations.}. Although our results do not match the observations perfectly, the cosmography simulated using our constraints captures the essence - in terms of mass and distances - of the observed Local Group.

\begin{table}
\begin{center}
 \begin{tabular}{l l l l}
Property &Simulated LG  & Observed LG & Reference\\
   \hline
   \hline
$M_{\rm MW}$  & $6.57\times10^{11} M_{\odot}$ & $ 10^{12} M_{\odot} $& [1,2,3] \\
$M_{\rm M31}$  & $8.17\times10^{11} M_{\odot}$ & $8.2\times10^{11} M_{\odot}$ & [4]\\
$M_{\rm M33}$  & $2.02\times10^{11} M_{\odot}$ & $6\times10^{10} M_{\odot}$ & [5] \\
$r_{\rm MW}$  & $220~\rm kpc$ & $253~\rm kpc $& \\
$r_{\rm M31}$  & $245~\rm kpc$ & $237~\rm kpc $ &\\
$r_{\rm M33}$  & $183~\rm kpc$ & $100~\rm kpc$ & \\

   \hline

 \end{tabular}
 \end{center}
\caption{The $z=0$ properties of the simulated and observed Local Group. From the top row own, we show the following properties: the mass of the MW's halo ($M_{\rm MW}$), the mass of M31's halo ($M_{\rm M31}$), the mass of M33's halo ($M_{\rm M33}$), the virial radius of the MW halo ($r_{\rm MW}$), the virial radius of M31's halo ($r_{\rm M31}$), and the virial radius of M33's halo ($r_{\rm M33}$). Note that the ``observed'' virial radii are calculated from the observed virial masses; they are thus unreferenced. The references are as follows: [1]~\citet{2008ApJ...684.1143X}; [2]~\citet{2002ApJ...573..597K}; [3]~\citet{2007MNRAS.379..755S}; [4]~\citet{2008MNRAS.389.1911S};[5]~\citet{2003MNRAS.342..199C}}
\label{table:cosm}
\end{table}

In addition to our gas dynamical SPH simulation, we also have a DM only version seeded from the same initial conditions. A comparison between the two simulations has already been highlighted in \cite{2010MNRAS.401.1889L} and \cite{2010MNRAS.405.1119K}. The DM only simulation has similar spatial and mass resolution and reproduces the same three main haloes as the gas-dynamical simulation. We use the DM only simulation solely in Section~\ref{sec:mimic}, where we try to find a recipe by which particles in a DM only simulation can be used to reproduce the radial distribution of the stellar halo, without the necessity of a complicated semi-analytical model.

\subsection{The halo and subhalo finding algorithm}\label{sec:AHF}
In this section, we explain how our halo and subhalo finding algorithm works.
In order to identify haloes and subhaloes in our simulation we have run the
MPI+OpenMP hybrid halo finder \texttt{AHF} (\texttt{AMIGA} halo finder, to be
downloaded freely from \texttt{http://popia.ft.uam.es/AMIGA}) described in
detail in \cite{2009ApJS..182..608K}. \texttt{AHF} is an improvement of the
\texttt{MHF} halo finder \citep{2004MNRAS.351..399G}, which locates local
over-densities in an adaptively smoothed density field as prospective halo
centres. The local potential minima are computed for each of these density
peaks and the gravitationally bound particles are determined. Only peaks with
at least 20 bound particles are considered as haloes and retained for further
analysis. In practice for this work, we only consider subhaloes with more than 100 particles. We would like to stress that our halo finding algorithm automatically
identifies haloes, sub-haloes, sub-subhaloes, etc. For more details on the
mode of operation and actual functionality we refer the reader to the
code description paper \citep{2009ApJS..182..608K}. 

For each halo, we compute the virial radius $r_{\rm vir}$, that is the radius
$r$ at which the density $M(<r)/(4\pi r^3/3)$ drops below $\Delta_{\rm
  vir}\rho_{\rm back}$. Here $\rho_{\rm back}$ is the cosmological background
matter density. The threshold $\Delta_{\rm vir}$ is computed using the spherical
top-hat collapse model and is a function of both cosmological model and
time. For the cosmology that we are using, $\Delta_{\rm vir}=355$ at $z=0$.  

Subhaloes are defined as haloes which lie within the virial radius of a more
massive halo, the so-called host halo. As subhaloes are embedded within the
density of their respective host halo, their own density profile usually
shows a characteristic upturn at a radius $r_t \lsim r_{\rm vir}$, where
$r_{\rm vir}$ would be their actual (virial) radius if they were found in
isolation.\footnote{Please note that the actual density profile of subhaloes
  after the removal of the host's background drops faster than for isolated
  haloes \citep[e.g.][]{2004ApJ...608..663K}; only when measured within the
  background still present will we find the characteristic upturn used here to
  define the truncation radius $r_t$.}  We use this ``truncation radius''
$r_t$ as the outer edge of the subhalo and hence subhalo properties
(i.e. mass, density profile, velocity dispersion, rotation curve) are
calculated using the gravitationally bound particles inside the truncation
radius $r_t$. For a host halo we calculate properties using the virial radius $r_{\rm vir}$.

We build merger trees by cross-correlating haloes in consecutive simulation
outputs. For this purpose, we use a tool that comes with the \texttt{AHF}
package and is called \texttt{MergerTree}. As the name suggests, it serves the
purpose of identifying corresponding objects in the same simulation at different
redshifts. We follow each halo (either host or subhalo) identified at redshift
$z=0$ backwards in time, identifying as the main progenitor (at the previous
redshift) the halo that both shares the most particles with the present halo \textit{and} is
closest in mass. The latter criterion is important for subhaloes given that all their
particles are also typically bound to the host halo, which is typically orders
of magnitude more massive. Given the capabilities of our halo
finder \texttt{AHF} and the appropriate construction of a merger tree,
subhaloes will be followed correctly along their orbits within the
environment of their respective host until the point where they either are
tidally destroyed or directly merge with the host.

\subsection{Identifying the stellar halo}
In this section we explain the nomenclature used for the different particle sets in our analysis. We exclude from our analysis the inner baryonic component (i.e. the galactic disc) and use the term ``outer halo'' to refer to the region between $0.1~\rvir$ and $\rvir$. We identifying those DM and stellar particles that at $z=0$ are within this region and then excise all particles bound to substructures. We call the ``swiss cheese'' like remains the \textit{diffuse} (stellar or DM) halo. We focus our analysis on the origin of this diffuse component.

We refer to star particles in the diffuse outer halo as the ``stellar halo''. For each of these particles, our simulation provides us with the age of the universe when it formed. Since star particles by construction can only be formed in high density environments (and are thus bound to a subhalo at the moment of their birth) we can use the star particle's age to locate the appropriate snapshot and thus the (sub)halo in which the star was spawned. If the star formed in the main progenitor of its $z=0$ host, we say the particle formed ``in-situ''; if it formed in any other halo we say it formed ``ex-situ''. 

Ex-situ stellar halo particles have thus been stripped from the accreted substructure within which they were born and orbit within the parent as debris material. We thus also refer to these particles as ``stripped'' star particles and use this term interchangeably with ``ex-situ''. In situ halo star particles were formed in the centre of the progenitor of the host halo and are then pushed out by merging or migrational processes \citep[e.g.][]{2009ApJ...702.1058Z}. 

For the diffuse DM halo, particles may also be stripped from accreted subhaloes. Unlike star particles however DM particles may also be ``smoothly'' accreted, in other words accreted by the main progenitor either individually from the ambient cosmic background or in substructures below our subhalo resolution limit of 100 particles.

In this work we have used the term “stripped” particles to denote those particles that become unbound from the subhalo in which they were accreted by any physical mechanism. Our term is general in that we do not differentiate between tidal or resonant stripping \citep{2009Natur.460..605D}.

Throughout this paper we choose not to stack our three haloes into single plots, since the variety of results obtained is significant and due to the different and unique merger histories of each halo.

\section{Results}
\label{sec:results}

\begin{figure*}
\includegraphics[width=40pc]{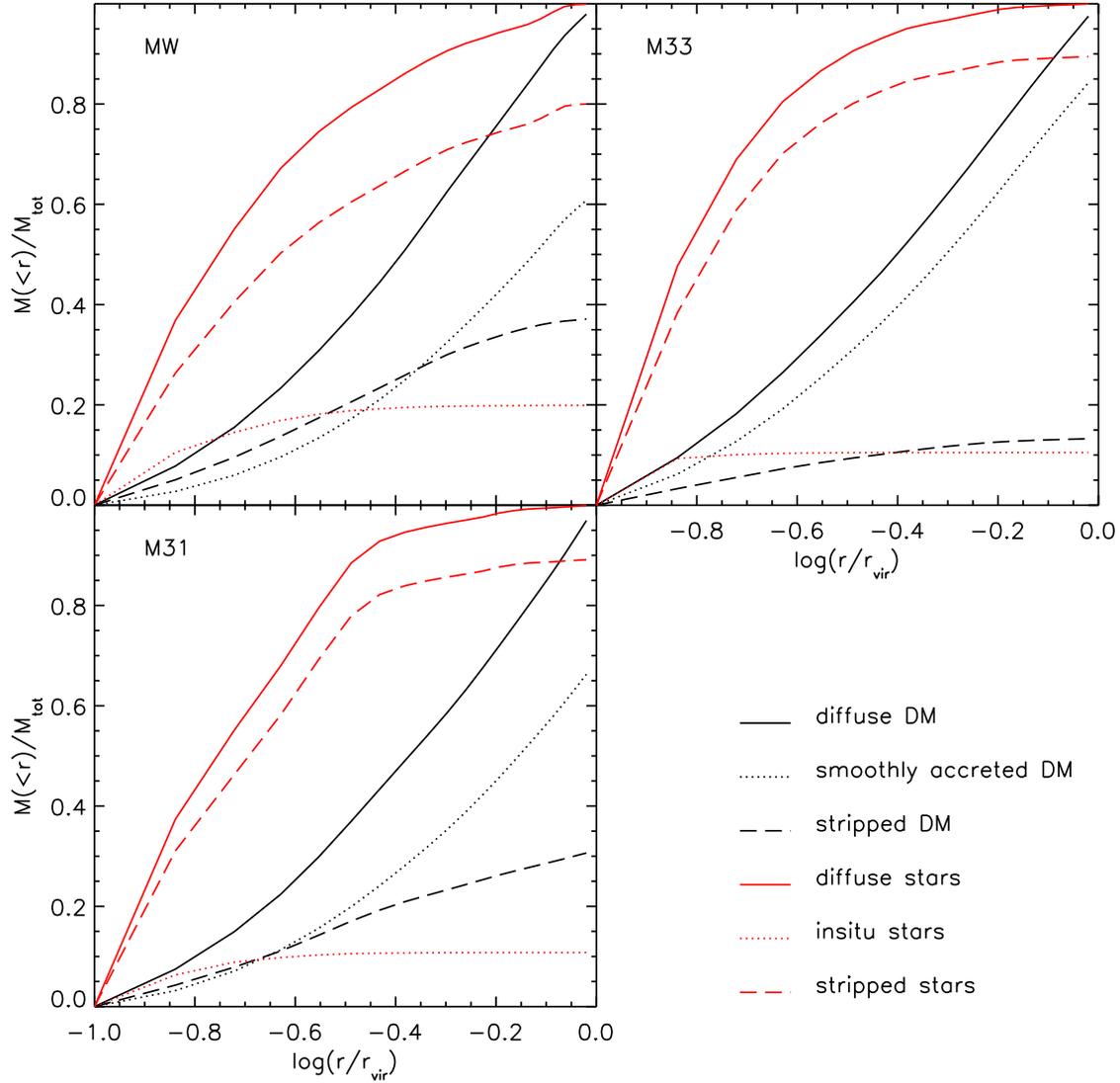}
\caption{The cumulative mass profile (i.e. total mass fraction within a radius $r$) for the diffuse component of our MW (upper left), M31 (lower left) and M33 (upper right) outer haloes. Both DM (black) and stellar (red) curves are normalized by the total mass in the respective diffuse component within the outer halo. The solid red curve shows the mass profile of the diffuse stellar component. The red dot-dashed line shows the mass profile for those stars born within the main progenitor, known as ``in-situ'' stars. The dashed red line shows the mass profile for stars accreted in clumps and later stripped from them such that at $z=0$ they are bound just to the main halo. Similarly, the solid black curve shows the mass profile of the diffuse DM component. The dot dashed black curve shows this quantity for smoothly accreted DM and the  dashed black line represents the mass profile for stripped debris material. }

\label{fig:radprof}
\end{figure*}

We begin by studying the radial distribution of mass within each (DM and stellar) component. In Fig.~\ref{fig:radprof} we show the diffuse mass interior to a given radius as a function of radial distance from the centre, for DM (black solid) and stars (red solid) for our three galaxy haloes. We normalise each component by the total diffuse halo mass within the outer halo. Note that the diffuse DM makes up roughly 80\% of the halo's full virial mass \citep{2004MNRAS.352L...1G} and is used in the semi-analytical investigations of  \cite{2010MNRAS.406..744C} as a proxy with which to study the stellar halo. 

When comparing the stellar component to the DM component (the solid red line to the solid black line), a very stark difference is immediately visible. Although star and DM particles are treated equally in the simulation´s gravity calculation, their $z=0$ distribution differs dramatically in that the star particles are highly concentrated towards the centre of the outer halo while the DM roughly follows an NFW distribution \citep[not shown here, but see e.g.][]{2010arXiv1008.5114W}.

The extreme central concentration of stars is likely due to radiative cooling, which causes gas particles to lose energy, fall to the centre of the halo where the densities are high enough for star formation, adiabatically contract the DM and deepen the potential, a result that has been know since at least \cite{1986ApJ...301...27B}. Yet star particles can also fall to the centre by losing angular momentum through dynamical friction against the halo background (as can the DM) a process known to be more effective for clumps that contain baryons \citep{2001ApJ...560..636E,2008ApJ...685L.105R}. These two routes of galaxy formation are both important and have been studied in great detail by e.g. \cite{2010ApJ...725.2312O}. Thus in order to better understand how this difference in radial profile of the diffuse component has arisen, we must examine in greater detail each component individually.

We now focus on the stellar component, specifically on its two constituent subsets: in-situ and ex-situ star particles. The stripped and in-situ stars display considerably different radial profiles (red dotted and red dashed curves respectively in Fig.~\ref{fig:radprof}), perhaps not surprising given their different physical origins. Whereas the cumulative in-situ stellar mass rarely grows beyond $0.2~\rvir$, the stripped stars are found in abundance at all radii throughout the stellar halo.

Note that in-situ stars contribute very little to the total mass when averaged over the entire diffuse halo. The outer haloes are dominated by ex-situ star particles that comprise $\sim 80$\% of the MW's outer halo and $\sim 90$\% of M31's and M33's outer halo. The exact amount is due to the unique merger history of each individual halo. Because the outer halo is mostly made up of these stripped particles, their mass profiles dominate the total mass profile of the outer halo. The fact that the stellar halo is composed primarily of stripped stars bodes well since this population has direct counterpart in the DM component.

\begin{figure*}
\includegraphics[width=40pc]{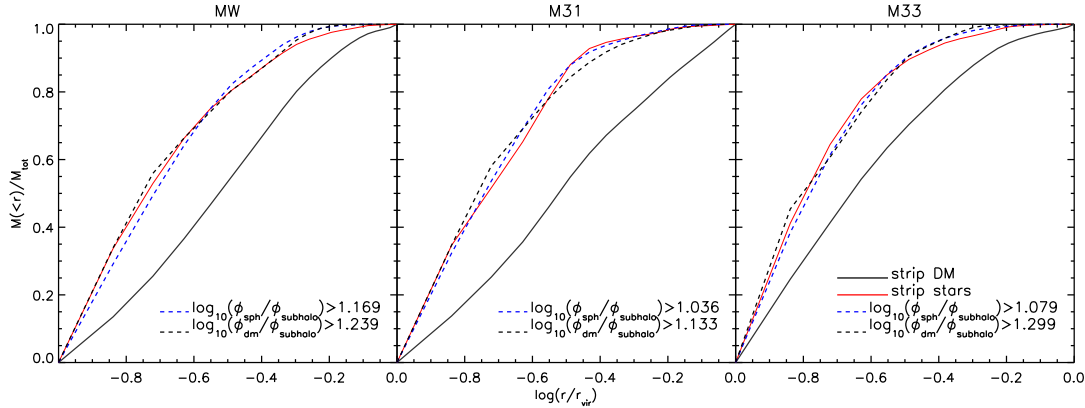}
\caption{The cumulative mass profile (i.e. total mass fraction within a radius $r$) for just the stripped component of the diffuse haloes of our MW (left), M31 (centre) and M33 (right) haloes. Both DM (black) and stellar (red) curves are normalized by the total mass in the respective stripped component within $\rvir$. The dashed blue curve corresponds to those DM particles in the gas-dynamical simulation whose potential at accretion ($\phi_{\rm sph}$) was greater than $10^{1.169}$, $10^{1.036}$, $10^{1.079}$ times that of the host ($\phi_{\rm subhalo}$) they were accreted in. The dashed black line corresponds to those DM particles in the DM only simulation whose potential at accretion ($\phi_{\rm DM}$) was greater than $10^{1.239}$, $10^{1.133}$,  and $10^{1.299}$ times that of the host for the MW, M31, and M33 respectively. See section~\ref{sec:mimic} for more on how these thresholds were obtained.}
\label{fig:stripped}
\end{figure*}

We now look at the two components of the diffuse DM halo, the smoothly accreted and stripped particles. Note that the smoothly accreted component has no stellar counterpart in since a negligible number of star particles are ``pre-stripped'' and smoothly accreted. From Fig.~\ref{fig:radprof} we see that the smoothly accreted DM constitutes the major part of the outer diffuse DM halo, contributing $\sim 60 - 70\%$ to its mass. Furthermore, its radial distribution has roughly the same shape as the total outer DM halo (not surprisingly since it dominates the halo's mass) and is markedly different from the stripped DM debris.

Unlike the smoothly accreted DM, the stripped DM (shown in Fig.~\ref{fig:radprof} as the dashed black line), has a direct counterpart among the star particles (see above). Yet the two stripped profiles have completely divergent shapes and it is difficult to find any similarities between the two populations. Furthermore, stripped DM contributes just $\sim40\%$, $\sim30\%$ and $\sim 25\%$ to the total mass of the outer halo for the MW, M31 and M33 respectively (versus $\sim 80-90\%$ for stripped stars).

But how different are the radial profiles of just the stripped stellar and DM components when normalized to their respective $r_{\rm vir}$ masses? In Fig.~\ref{fig:stripped} we plot the radial profiles of these two components. The two curves clearly deviate substantially from each other, both qualitatively and quantitatively. Stripped stars (red lines of Fig.~\ref{fig:stripped}) are, as noted earlier, significantly more centrally concentrated then the stripped DM (black line). Note that the cause for this discrepancy cannot be the deeper potential of the host - adiabatic contraction of the DM due to radiative cooling of the gas only affects the region interior to $0.1r_{\rm vir}$ \citep[see Fig.2 of][]{2010MNRAS.401.1889L}

Thus despite the fact that these two populations are born out of the same physical processes (and treated equally by the simulations gravity calculator), they still have at $z=0$ very diﬀerent radial distributions. We now examine the origin for this dichotomy.

\subsection{Examination of the potential at accretion}
\label{sec:potential}

The chance that a particle will at some point become stripped from the clump that it was bound to at accretion, depends on a variety of properties of the sub-clump (for example its orbit, internal structure, spin, etc). One important factor is the nature of the potential well of the subhalo that each accreted particle sits in at accretion. The potential of each particle at accretion can be calculated by assuming that the accreted subhalo obeys spherical symmetry. The potential at a distance $r$ from the subhalo's centre is thus:
\begin{equation}
\phi(r)=G\int_{0}^{r}\frac{M(<r')}{r'^{2}}dr' +\phi(0)
\end{equation}
where $\phi(0)$ is normalized such that the potential is null at infinity \citep[cf the appendix of][]{2009ApJS..182..608K}. At the subhaloes edge the potential is
\begin{equation}
\phi_{\rm subhalo} = -\frac{GM_{\rm tot}}{r_{\rm vir}}
\end{equation}

Since the depth of a potential well at the centre of a subhalo is a measure of the concentration \citep[which in turn depends on other subhalo dependent properties like the total mass or formation time - see e.g. ][among others]{2003MNRAS.339...12Z,2011MNRAS.411..584M} we normalize the potential of each particle by the potential at the subhaloes edge to obtain a dimensionless number that is more or less independent of \textbf{global} subhalo properties.

\begin{figure*}
\includegraphics[width=40pc]{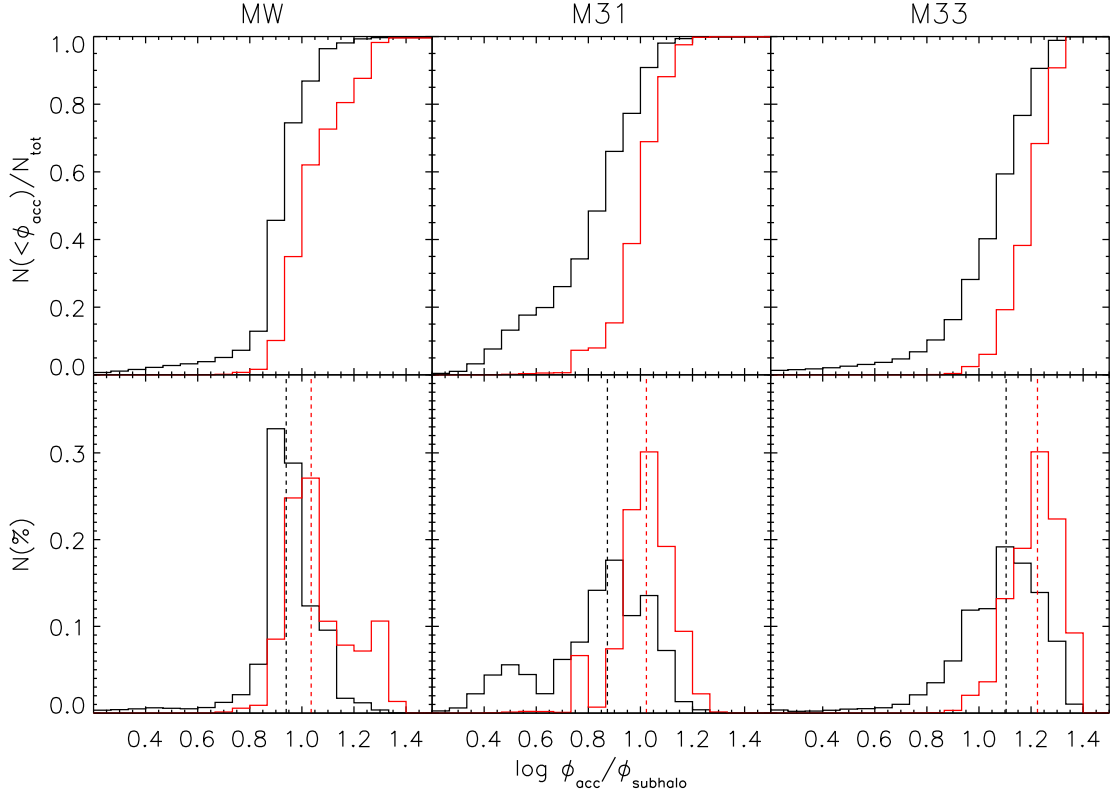}
\caption{The cumulative (top) and differential (bottom) distribution of the logarithm of a particle's potential at accretion ($\phi_{\rm acc}$) normalized to the value of the potential at the subhalo's virial radius ($\phi_{\rm subhalo}$). In black we show this distribution for DM and in red for star particles. The median values are shown by the vertical dashed lines and are $10^{1.03}$ ($10^{0.93}$), $10^{1.02}$ ($10^{0.87}$), and $10^{1.22}$ ($10^{1.1}$), for the stars (DM) in the MW, M31 and M33 haloes.}
\label{fig:phi}
\end{figure*}

We plot the distribution of the logarithm of this value in Fig.~\ref{fig:phi} for stars (red) and DM (black). The vertical dashed lines indicate the median values, which for the stars are always more than 10$\times \phi_{\rm subhalo}$. This means that the stars that get stripped and which subsequently end up dominating the stellar halo had, at the time of accretion, a potential ten times greater then that at the edge of their host subhalo. This is a reflection of the fact that stars are ``more bound'' than other particles in the same subhalo; yet there is some tension because the more bound a particle is to its subhalo, the less likely it is to be stripped from it. The stripped DM, on the other hand, occupies ``less bound'' regions of the subhalo's potential at accretion, evident by the fact that the median potential is always significantly lower than that for the star particles. It may be counter-intuitive that so many stripped particles have apparently high values of $\log\phi_{\rm acc}/\phi_{\rm subhalo}$, but this is a direct result of the fact that particles in subhaloes are not uniformly distributed; instead the majority of the particles that make up a subhalo are concentrated in the centre of the subhalo, where the potential is much higher than at the subhalo's edge.

In essence, Fig.~\ref{fig:phi} reflects the distribution of binding energies at accretion of stripped particles. We could have plotted this value instead, but by normalizing by the potential at the halo's edge we obtain a dimensionless quantity which is subhalo independent.

Fig.~\ref{fig:phi} reveals that DM particles are stripped from the fluffier outer regions of a subhalo, while star particles are stripped from the deep interior of their subhaloes. This can explain why the number of stripped star and DM particles is so different: not only are there fewer stars to strip to begin with, but they are harder to strip after accretion.

\begin{figure}
\includegraphics[width=20pc]{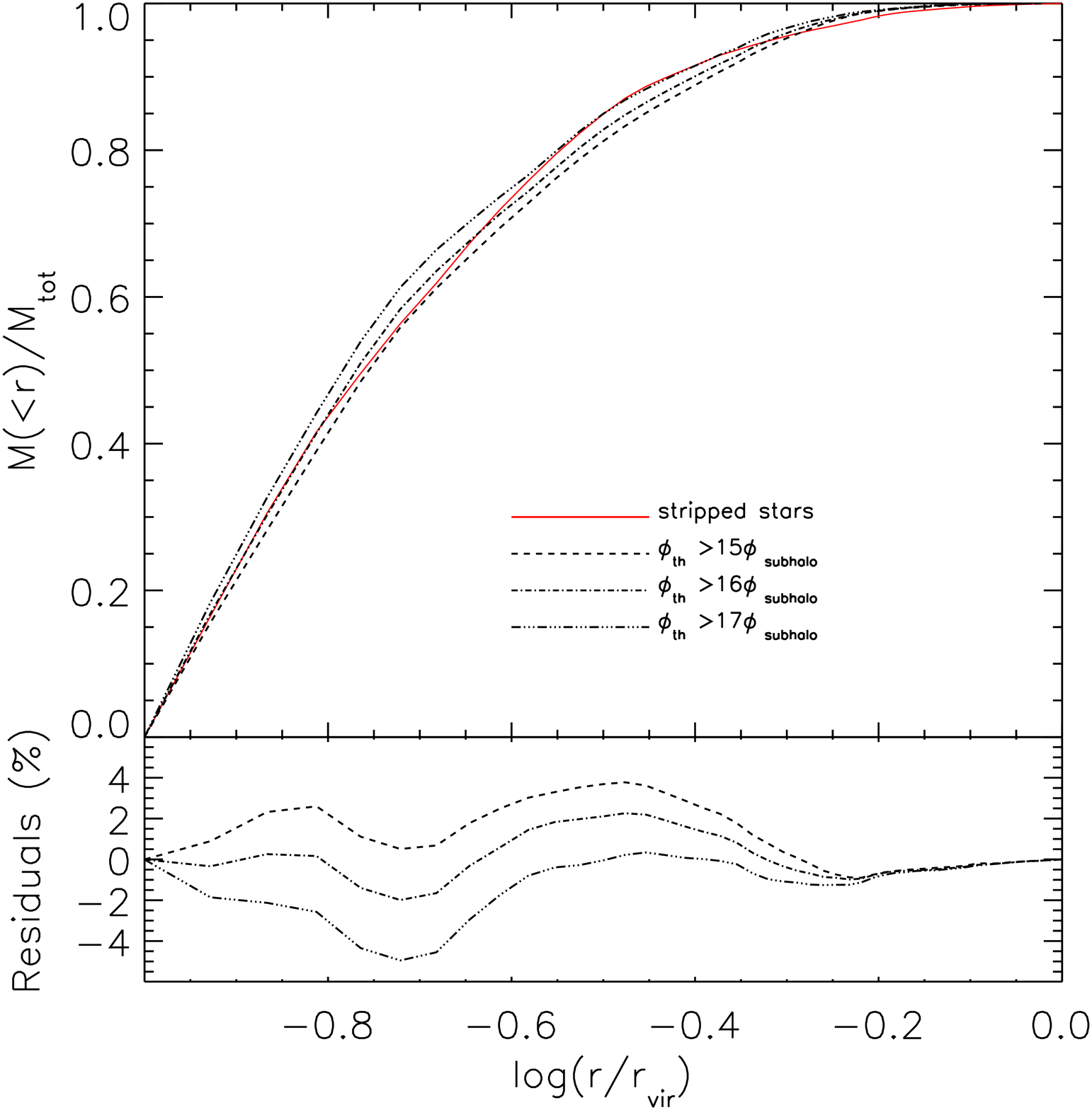}
\caption{The co-added cumulative diffuse halo radial profile at $z=0$ averaged over our 3 haloes. In red we show the stellar distribution. The black lines indicate different thresholds in the potential, above which DM particles in our DM only simulation were selected. The dashed line is thus the radila distribution of DM particles who at accretion had potentials greater than $15\phi_{\rm subhalo}$ times the potential at the subhaloes edge. The dot - dashed line is for potentials greater than $16\phi_{\rm subhalo}$  and the triple dot dashed line is for potentials greater than $17\phi_{\rm subhalo}$. The bottom panels shows the residuals when the stellar distribution is subtracted.}
\label{fig:coadded}
\end{figure}

\subsection{Finding a DM subset that mimics the stellar halo}
\label{sec:mimic}

We wish to thus identify a subset of the DM that follows at $z=0$ the same radial distribution as the stripped stars but that does not - in any way - depend on stellar properties. In this way, pure DM-only simulations may be used and a subsample that accurately reproduces the stellar radial distribution can be obtained.

From Fig.~\ref{fig:phi} we know that stars were more tightly bound in bigger subhaloes at the moment of accretion, while the stripped DM was more loosely bound to smaller subhaloes when they fell in. We thus wish to choose DM particles that sit at the same depth (or deeper) of the potential well as the stripped stars. In principle we could easily select a sub-sample of DM particles whose distribution of  $\log\phi_{\rm acc}/\phi_{\rm subhalo} $ perfectly matches that of the stars; yet we wish our criteria for sub-selecting the DM to be independent of any stellar properties.

We thus define a threshold $\phi_{\rm th}$ for the value of $\phi_{\rm acc}/\phi_{\rm subhalo}$ above which all DM particles are selected as a proxy for stars. After defining a threshold and obtaining a DM sub-sample, we examine the radial distribution of these particles and compare it (by calculating $\chi^{2}$) to the star particle curves in Fig.~\ref{fig:stripped}.  By smoothly varying $\phi_{\rm th}$ and examining the value of $\chi^{2}$ for each halo, we are able to obtain a DM sub-sample which very closely matches the stellar radial distribution.

Although the optimum $\phi_{\rm th}$ varies slightly across our three haloes, it does so weakly having values of $\phi_{\rm th~MW}=10^{1.169}\approx15$, $\phi_{\rm th~M31}=10^{1.036}\approx11$,  and $\phi_{\rm th~M33}=10^{1.079}\approx12$,  for the MW, M31 and M33 respectively. We plot the radial distribution of DM particles that meet the $\phi_{\rm th}$ criteria in Fig.~\ref{fig:stripped} as the dotted blue line, and note that by selecting these subsets we nearly perfectly recover the radial stellar distribution for each halo. Note that roughly the same number of DM particles meet these criteria as halo star particles.

Yet in order to develop a rule by which DM only simulations can be used to study the stellar halo, we must attempt to match the radial distribution of a subset of DM particles in a DM only simulation to the stripped star particles in the gas-dynamical one. The thresholds mentioned above work well for the  gas-dynamical simulation but, since the potential of each halo is aﬀected by adiabatic contraction due to the collapsed baryons, the limit is not directly comparable to DM only simulations.

In order to address this concern, we perform the exact same analysis we have thus far presented on our DM only simulation. We perform the same $\chi^{2}$ minimization test to obtain the best fit radial profile of DM particles in the DM only simulation. We show this distribution as the dashed black line in Fig.~\ref{fig:stripped}. In order to reproduce the stripped stellar distribution we need to select DM particles that are slightly deeper than their counterparts in the gas dynamical simulation, having potentials greater than $\phi_{\rm th~MW}=10^{1.239}\approx17$, $\phi_{\rm th~M31}=10^{1.133}\approx14$,  and $\phi_{\rm th~M33}=10^{1.299}\approx20$,  for the MW, M31 and M33 respectively. The necessity of selecting particles at a deeper part of the potential in DM only simulations is due to the relatively shallower potentials in simulations without baryons.

We note that the number of stripped DM halo particles that meet the $\phi_{\rm subhalo}$ criteria in the gas dynamical simulation is: $\sim 2\%$, $5\%$ and $6\%$ of the total stripped halo for the MW, M31 and M33 respectively. This fraction is roughly the same as is found in the DM only simulations where $\sim 1\%$, $3\%$ and  $3\%$ of the diffuse stripped halo meet the criteria for the MW, M31, and M33 respectively. Furthermore the absolute number of particles that meet these criteria is roughly the same across all three haloes and is the same order of magnitude as the number of stars in the stripped stellar halo.

We now attempt to obtain a universal threshold in the potential at accretion of DM particles in DM only simulations that reproduce the stripped stellar profile at $z=0$. We begin by averaging the stripped stellar profile of our three haloes and plotting it as the red line in Fig.~\ref{fig:coadded}. We then co-add the diffuse stripped DM halo particles and select DM particles according to whether they are above or below the threshold potential $\phi_{\rm th}$ at accretion. We examine three fiducial values for the threshold: $\phi_{\rm th}=15\phi_{\rm subhalo}$, $16\phi_{\rm subhalo}$ and $17\phi_{\rm subhalo}$, and plot the radial distribution of DM particles that meet these criteria as the dashed, dot dashed and triple dot dashed lines in Fig.~\ref{fig:coadded}. In the bottom panel of Fig.~\ref{fig:coadded} we show the residual value when the DM profile is subtracted from the stellar profile. We note that the best value is $\phi_{\rm th} =16\phi_{\rm subhalo}$, and returns a distribution that is within 2\% of the stellar profile. Thus, a rule as simple as selecting particles according to their potential at accretion is able to reproduce the radial distribution resulting from the complicated physics of star formation.

Note that increasing the threshold, selects particles from deeper in the potential well (e.g. $\phi_{\rm th}=17\phi_{\rm subhalo}$, the triple dot dashed line in Fig.~\ref{fig:coadded}) of the halo they were accreted in and results in a more centrally concentrated $z=0$ distribution. This is because particles with higher potentials at accretion are harder to strip - this requires their subhaloes to be on highly radial orbits such that pericentric passages bring them into regions where tidal forces are strong enough to rip them from their hosts. This occurs only towards the centre of the halo and results in their deposition closer to the halo centre.

We have now obtained a ``rule'' by which DM particles can be selected in order to reproduce the $z=0$ radial distribution of star particles. The rule is fairly simple: for each DM particle that is bound to a subhalo at accretion and later stripped, it must be sitting, at the moment of accretion, at a position in the potential well of its host that is deeper then $\sim 12~\phi_{\rm subhalo}$ in a radiative gas-dynamical simulation and at a potential that is deeper than  $\sim 16~\phi_{\rm subhalo}$ in a DM only simulation. In this way a DM only simulation can be used to study the stellar halo without a complicated semi-analytical model to treat the baryons. This rule is consistent across our three haloes.

Note that despite this universality,  our three halos have very different histories. The halo of M33 had a relatively quiet mass accretion history, growing by smooth accretion and minor mergers for the past $\sim10$~Gyrs. In contrast, the haloes of M31 and the MW experienced a more violent past with major mergers occurring more frequently and more recently. Our small sample size of just three haloes thus represents a very wide variety of mass accretion histories. Although it is difficult to derive a universal relation based on a sample size of just three haloes, the fact that we find the exact same threshold value of $16\phi_{\rm subhalo}$ across haloes with very different mass accretion history hints at the possibility that this is indeed a universal relation.

\subsection{The assembly of the stripped halo}
\begin{figure*}
\includegraphics[width=40pc]{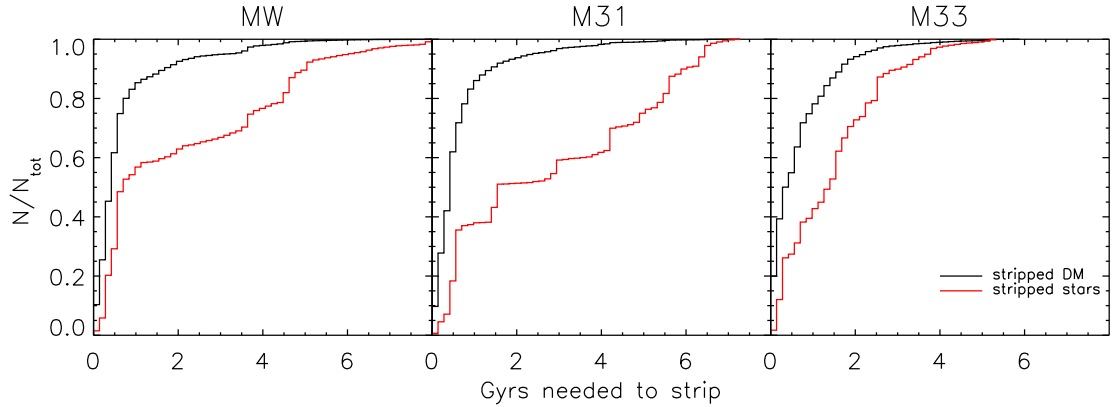}
\caption{The fraction of particles that are stripped from their subhaloes after a given period of time. In red we show star particles and in black, DM. The fact that at accretion star particles are deeper in their host potential is reflected by the fact that it takes longer to strip a given fraction of star particles than DM ones.}
\label{fig:assembly}
\end{figure*}

In Section~\ref{sec:potential} we showed that star particles and DM particles occupy different parts of a (sub)halo's potential at accretion. Since star particles sit deeper in their host's potential we can infer that they are thus harder to strip and will thus become unbound later than DM. For each particle in the stripped halo, we can thus measure how many Gyrs after accretion the particle becomes unbound from its substructure and begins to orbit in the diffuse halo as debris. In Fig.~\ref{fig:assembly} we show the cumulative distribution of this quantity. Star particles take a considerably longer amount of time to become unbound than their DM counterparts. For example, if one examines the fate of particles $\sim1$~Gyr after accretion, we see that $\sim70\%$ of DM particles have been stripped while just $\sim 40\%$ of star particles have been stripped. Some star particles are so deep in the potential wells of their hosts, that it can take up to half a Hubble time to strip them, while a very small fraction of DM particles that end up getting stripped will still be bound after such a long period - they will be stripped much earlier.

\begin{figure*}
\includegraphics[width=40pc]{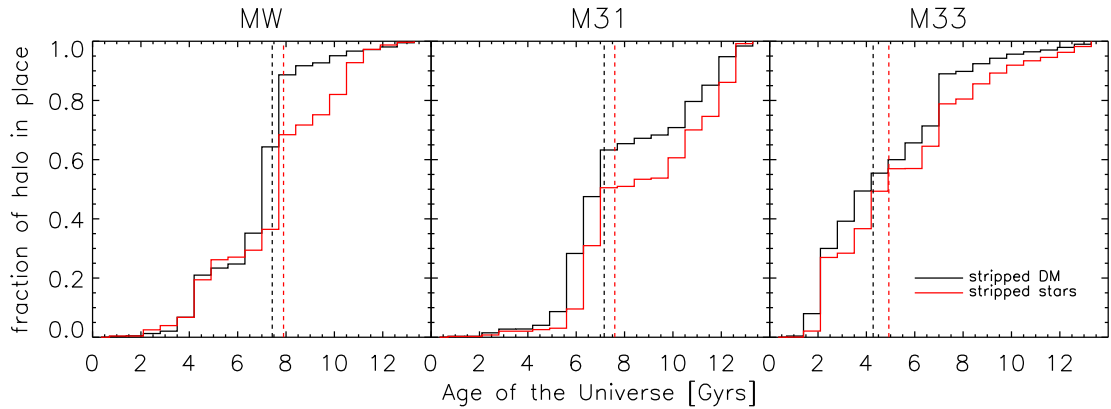}
\caption{The fraction of the stripped stellar (red) and DM (black) halo, thats in place as a function of the age of the Universe. The dashed lines show the median of the distribution: the age at which $50\%$ of the halo has been built up. The DM median is always earlier then the stellar median, reflective of the fact that at a given time, more of the DM halo is in place then the stellar halo. Half the mass of the stripped DM halo is in place  0.47~Gyrs, 0.44~Gyrs and 0.65~Gyrs before the stellar halo for the MW, M31 and M33 respectively. }
\label{fig:t_strip}
\end{figure*}

The fact that it takes star particles significantly longer to be stripped than DM particles leads to the conclusion that the stellar halo was assembled later than the diffuse DM halo. In Fig.~\ref{fig:t_strip} we show the fraction of the stripped stellar and DM halo in place as function of time since the Big Bang. Note that after $\sim5$~Gyrs, the DM halo grows faster that the stellar halo - reflective of the ease with which DM particles are stripped. As a result, the stripped stellar halo is considerably younger than the DM one. In Fig.~\ref{fig:t_strip} we also show the age at which 50\% of the halo debris has already been deposited and note that this is always earlier for the DM halo by around $\sim0.5$~Gyrs.

For each particle in our simulation we know the redshift at which it was accreted and the redshift at which it was stripped. We may thus ask the question: ``after accretion, what fraction of the future particle's life is spent in its subhalo and what fraction is spent orbiting the main halo as stripped debris''. This question gives us a feel for how efficient tidal stripped is in our subhaloes. DM particles spend \textit{on average} 10\% of their post-accretion life still bound to their subhalo of origin and massive 90\% of their post-accretion life orbiting as debris. As expected from the above arguments, stellar particles spend more time still bound to their subhaloes of origin: on average they spend 15\% of their post-accretion life in subhaloes and just 85\% of their time orbiting as debris. Note that these numbers are averages over all particles. Many particles - e.g. those accreted in substructures on circular orbits - will spend more time bound to their substructures then orbiting as debris material.

\section{Discussion and Conclusions}
\label{sec:conclusion}
In this paper we have studied three Galaxy sized haloes formed in a constrained simulation of the Local Group. The three galaxies have approximately the same size and relative positions (as well as some other properties) as the observed Local Group members and are therefore refereed to as MW, M31 and M33. By construction, the galaxies are formed in an environment whose bulk properties (e.g. distance to a Virgo mass cluster) closely match observations. We use our constrained Local Group to focus on the similarities and differences in the origin of the DM and stellar halo.

We have focused our analysis on attempting to understand why stripped DM particles orbiting in the parent halo, have a completely different radial distribution than ex-situ star particles, who were unbound by the exact same process. Specifically, the stars are more centrally concentrated then the DM, which roughly follows an NFW profile \citep[e.g.][]{2010arXiv1008.5114W}. Their different $z=0$ distribution implies that the process of tidal stripping must be operating on different subsets of the particles' distribution.

We searched for a subset of the DM particles that is indistinguishable from the stellar halo, in order to understand how these very similar components evolved differently from each other. We found that DM particles (in a gas-dynamical simulation) that had potentials at accretion of at least $\sim12$ times the potential at the subhalo's edge, are able to closely match the radial distribution of star particles. In a DM only simulation, the threshold above which selected DM particles reproduced the stellar halo distribution was $\sim 16$ times the potential at the subhaloes edge.

Our threshold value is constant across the three haloes we have studied with very small scatter, despite the fact that our small sample size of three (roughly) Milky Way sized haloes have very different mass accretion histories. 

We have thus found a method through which the stellar halo may be modelled without the need of running a semi-analytical model to treat the baryons. By simply selecting those particles at accretion whose potential is greater than the threshold value quoted here, a set of particles can be identified that nearly perfectly matches the stellar distribution.

The implications of this work are therefore that the stripped stellar halo reflects the fate of material that sits deep in a halo's potential well in an absolute sense, not in a relative one. It is important to note that selecting a fraction of DM particles according to their \textit{relative} binding energy at accretion (e.g. the 10\% most bound particles) will not successfully reproduce the $z=0$ stellar distribution  unless the 10\% most bound particles sit within a region of the potential which is greater than $\sim 16$ the potential at the host's edge. In fact selecting the 10\% most bound DM particles at accretion returns a halo profile which follows the DM and is around 20\% less centrally concentrated than the stars. This is because star formation occurs according to a local density criteria, not according to a ``global'' property, such as a particles binding energy relative to the entire subhalo. A given halo will only form stars in its centre if (a) it is large enough to retain its gas and shield it from photo-ionization \citep[e.g][]{2003ApJ...599...38B} and (b) the potential in its centre is deep enough such that the gas density may trigger star formation. Thus the likelihood of star formation depends just on structural parameters (like concentration) and whether densities in a halo's centre are high enough.

Additionally since the particles that sit in the deepest regions of the subhalo are harder to unbind, they will become unbound later - this implies that the stellar halo is younger then the DM halo: the star particles that compose the stellar halo were unbound from their substructures and deposited in the halo at lower $z$, than the DM. When we observe stars in the outer halo of the MW or M31, we must take care drawing conclusions regarding the DM haloes assembly.

The diffuse DM halo has profoundly different properties to the diffuse stellar halo. Its lack of central concentration dominates its global profile. Although there are some detailed differences among our three haloes in the relative contribution from in-situ and stripped material to the stellar background, the bottom line is that DM particles can not serve as a proxy for star particles unless care is taken in their selection. 

Two main results have been presented here. The first is that the stars that constitute the diffuse stellar halos form at the bottom of the potential well of subhaloes, and hence are more bound at infall than the corresponding DM particles that make the diffuse DM halo. This is a 'trivial' fact reproduced by all simulations of galaxy formation, yet there is no general consensus on the details underlying this fact. The novel and non-trivial other results is that one can find a simple mapping that enables the association of a subset of the DM particles with the stellar halo particles. The mapping is based on one single parameter, namely the scaled value of the depth of the potential well within which the halo stars formed. Such a simple mapping of the stellar halo and the subset of DM particles, is a robust outcome of the hierarchical nature of galaxy formation. Yet, the actual value of the parameter that controls the mapping most probably varies with the particular implementation of numerical simulations of galaxy formation. This parameter might vary with the details of how star formation and feedback processes are modeled as well as on the particular numerical schemes applied. This is posed here as an open question that we hope will be addressed by practitioners in the field.

\section*{Acknowledgments}
NIL is supported through a grant from the Deutsche Forschungs Gemeinschaft. AK is
supported by the MICINN through the Ramon y Cajal programme as well as the grants AYA 2009-13875-C03-02, AYA 2009-12792-C03, CSD2009-00064, and CAM S2009/ESP-1496. He further thanks Alan McGee for Creation Records. YH has been partially supported by the ISF (13/08). GY would like to thank the MICINN (Spain)  for financial support under project numbers FPA 2009-08958 and AYA 2009-13875-C03 and the SyeC Consolider project CSD 2007-0050. The simulations were performed and analyzed at the Leibniz Rechenzentrum Munich (LRZ), the Neumann Institute for Computing (NIC) Juelich and at the Barcelona Supercomputing Centre (BSC). We thank DEISA for  giving  us  access to  computing resources in these centres  through the DECI projects  SIMU-LU and SIMUGAL-LU. 

\bibliography{./ref}
\end{document}